\documentclass[12pt]{article}

\usepackage{bbm}
\usepackage{bbold}
\usepackage{color}
\usepackage{authblk}
\usepackage{latexsym}
\usepackage{epsfig,amssymb,euscript, mathrsfs,cite}
\usepackage{amsmath}
\usepackage{slashed}
\usepackage{mathrsfs} 
\usepackage{enumerate}
\definecolor{MyDarkBlue}{rgb}{0.15,0.15,0.45}
\usepackage[linktocpage=true]{hyperref}
\hypersetup{
colorlinks=true,
citecolor=MyDarkBlue,
linkcolor=MyDarkBlue,
urlcolor=MyDarkBlue,
pdfauthor={},
pdftitle={},
pdfsubject={hep-th}
}
\usepackage{chngcntr}
\counterwithin*{equation}{section}

\newcommand{\ex}{\mathrm{e}}
\newcommand{\ii}{\mathrm{i}}
\newcommand{\ext}{\widetilde{\mathrm{e}}}
\newcommand\diff{\mathrm{d}}

\newcommand{\vol}{\mathrm{vol}}

\newcommand{\R}{\mathbb{R}}
\newcommand{\C}{\mathbb{C}}
\newcommand{\CP}{\mathbb{CP}}
\newcommand{\Z}{\mathbb{Z}}
\newcommand{\N}{\mathbb{N}}

\newcommand{\ds}{\diff s^2}

\newcommand{\spindle}{\mathbb{\Sigma}}

\newcommand{\sigmag}{\Sigma_{\mathtt{g}}}

\newcommand{\ads}{\mathrm{AdS}}

\newcommand{\gammathreeE}{\tau^{(\mathrm{3d})}}

\newcommand{\gammasixL}{\gamma^{(\mathrm{6d})}}

\newcommand{\gammaeightL}{\gamma^{(\mathrm{8d})}}

\newcommand{\q}{q}
\newcommand{\rr}{r}

\newcommand{\es}[2] {\begin{equation} \label{#1} \begin{split} #2 \end{split} \end{equation}}

\begin{document}

\begin{titlepage}

\vskip 1cm

\begin{center}


{\Large \bf D6 branes wrapped on a spindle and $Y^{p,q}$ manifolds}

\vskip 1cm
{Pietro Ferrero}

\vskip 0.5cm

\textit{Simons Center for Geometry and Physics,\\
SUNY, Stony Brook, NY 11794, USA\\}

\end{center}

\vskip 0.5 cm

\begin{abstract}
\noindent  We present new solutions of 8d gauged supergravity which, upon uplift to type IIA, represent D6 branes wrapped on spindles. A further circle uplift gives 11d supergravity on a Calabi-Yau three-fold which is the cone over five-dimensional $Y^{p,q}$ manifolds. This highlights a connection between co-homogeneity one Sasaki-Einstein metrics in general dimension and the recently introduced spindle solutions in gauged supergravity. We find that a similar connection also exists for the small resolution of the Calabi-Yau cone over such manifolds. 

\end{abstract}

\end{titlepage}

\pagestyle{plain}
\setcounter{page}{1}
\newcounter{bean}
\baselineskip18pt

\renewcommand{\baselinestretch}{1.1}\normalsize
\tableofcontents
\renewcommand{\baselinestretch}{1.15}\normalsize
\newpage

\addtocontents{toc}{\protect\setcounter{tocdepth}{2}}

\section{Introduction}\label{sec:intro}

Among the several areas of intersection between string theory and mathematics, a particularly rich one is that of special holonomy manifolds, which naturally arise in compactifications of string theory and supergravity as a consequence of the existence of Killing spinors. Of special interest for us is the case of Sasaki-Einstein metrics\footnote{See \cite{Sparks:2010sn} for a review and references. }, of which only a finite number of examples were known until the work of \cite{Gauntlett:2004yd,Gauntlett:2004hh}. There, starting from a local metric which arises in the classification of supersymmetric AdS$_5$ vacua in 11d supergravity \cite{Gauntlett:2004zh}, the authors identified an infinite class of Sasaki-Einstein metrics in various dimensions, named $Y^{p,q}$. This success was followed by the discovery of an even larger class of Sasaki-Einstein metrics, called $L^{p,q,r_1,\ldots,r_{n-1}}$ in dimension $2n+1$ \cite{Cvetic:2005ft,Cvetic:2005vk}, obtained from the Wick-rotation of certain Kerr black holes.

More generally, supersymmetric solutions in supergravity theories can be classified assuming the existence of Killing spinors and studying the algebraic and differential conditions that this implies for the bilinear forms in such spinors \cite{Friedrich:2001nh,Gauntlett:2001ur,Gauntlett:2002fz,Gauntlett:2002sc}. This generally gives rise to geometries with $G$-structures, where $G$ is a Lie group, of which special holonomy manifolds are a special case where the torsion classes vanish. Over the years, supersymmetric solutions to gauged supergravity in various dimensions and with different amounts of Killing spinors have been classified, revealing interesting geometrical structures which, through the AdS/CFT correspondence, provide insights on strongly coupled quantum field theories. A particularly interesting example for this paper is that of so-called Gauntlett-Kim (GK) geometries \cite{Kim:2005ez,Kim:2006qu,Gauntlett:2007ts}, which arise in the classification of $\ads_2\times Y_9$ solutions of 11d supergravity and $\ads_3\times Y_7$ solutions of type IIB supergravity. Some of the simplest examples of GK geometries where the metrics are explicitly known have been recently revisited and interpreted as describing a situation in which branes are wrapped on two-dimensional orbifolds known as {\it spindles}, originally found for D3 branes \cite{Ferrero:2020laf}. Ever since, solutions describing branes wrapping spindles have been found in a large variety of examples \cite{Ferrero:2020laf,Hosseini:2021fge,Boido:2021szx,Ferrero:2021etw,Ferrero:2020twa,Cassani:2021dwa,Ferrero:2021ovq,Couzens:2021cpk,Ferrero:2021wvk,Giri:2021xta,Faedo:2021nub,Couzens:2022yiv,Couzens:2022aki,Arav:2022lzo,Suh:2022pkg,Suh:2023xse,Hristov:2023rel,Amariti:2023gcx,Amariti:2023mpg} and generalized to other two- \cite{Bah:2021mzw,Couzens:2021tnv,Suh:2021ifj,Suh:2021hef,Couzens:2021rlk,Suh:2021aik,Bah:2021hei,Couzens:2022yjl,Karndumri:2022wpu,Couzens:2023kyf} and higher-dimensional \cite{Faedo:2022rqx,Suh:2022olh,Couzens:2022lvg,Cheung:2022ilc,Bomans:2023ouw,Faedo:2024upq,Macpherson:2024frt} orbifolds.

All of the spindle solutions found so far in the literature are associated with dual conformal field theories and as such contain explicit AdS factors in the metric. However, this is not a necessary conditions and branes can be wrapped on spindles even in the absence of conformal invariance. We show this for all D$p$ branes in the companion paper \cite{Boisvert:2024jrl}. Here we focus in particular on the case of D6 branes, which reveals an interesting and previously unnoticed connection between $Y^{p,q}$ Sasaki-Einstein metrics and spindle solutions. To understand this, it is useful to recall that the case of D6 branes wrapped on a two-sphere was analyzed in 8d gauged supergravity in \cite{Edelstein:2001pu}, where it was shown that its uplift to eleven dimensions gives the conifold (or the small resolution thereof, according to the type of solution that one considers). Here we show that, similarly, the uplift of a solution describing D6 branes wrapped on a spindle gives 11d supergravity on the Calabi-Yau cone over a $Y^{p,q}$ manifold of dimension five. Starting from this observation, we show that it is possible to write the $Y^{p,q}$ Sasaki-Einstein metrics in general dimension $2n+1$ as $U(1)$ fibrations over a base given by the warped product betwen a spindle and a K\"ahler-Einstein space of complex dimension $n-1$, with metric of constant curvature. Note that this differs from the canonical way of writing these metrics, as the base space of the $U(1)$ fibration described above is {\it not} K\"ahler-Einstein itself. 

The fact that quotienting a Sasaki-Einstein metric along the Reeb vector in a suitable way leads to a K\"ahler orbifold metric in one dimension lower is not a new fact -- see {\it e.g.} \cite{Boyer:1999ms}. However, this procedure generically breaks supersymmetry, so what is non-trivial here is that we have {\it supersymmetric} solutions in 8d gauged supergravity which uplift to eleven-dimensional supergravity on the cone over five-dimensional $Y^{p,q}$ manifolds. More generally, for $Y^{p,q}$ manifolds of any dimension we are able to write the associated Killing spinors in a way that manifests supersymmetry on the spindle, thus showing that the reduction we propose always preserves supersymmetry. The two mechanisms for supersymmetry preservation on a spindle were discussed in \cite{Ferrero:2021etw}, where they are referred to as {\it twist} and {\it anti-twist}. We find that our point of view on $Y^{p,q}$ metrics is always associated with the twist condition, which is globally equivalent to the partial topological twist of \cite{Witten:1988ze} and allows for the spindle to be embedded as a holomorphic curve inside a Calabi-Yau manifold, as opposed to the anti-twist. Other examples of reductions of Sasaki-Einstein metrics that lead to supersymmetric K\"ahler orbiolds can be found in Section 5 of \cite{Gauntlett:2004zh}, Appendix B of \cite{Bianchi:2021uhn} and \cite{Macpherson:2024frt} for metrics on $\mathbb{WCP}^2$. See also \cite{Martelli:2023oqk,Inglese:2023tyc} for a discussion of the relationship between three-dimensional lens spaces and spindles.

In addition, we observe that in the case of D6 branes wrapped on $S^2$ there is a particularly simple supersymmetric solution in 8d gauged supergravity which uplifts to the conifold, while a slightly more complicated solution gives the small resolution of the conifold \cite{Edelstein:2001pu}. Similarly, we show that for spindles a more complicated solution than the one discussed above exists, which depends non-trivially on two coordinates in eight dimensions, whose uplift to 11d gives the small resolution of the Calabi-Yau cone over $Y^{p,q}$ -- a metric that was first found in \cite{Martelli:2007pv,Martelli:2007mk}.

The rest of this paper is organized as follows. In Section \ref{sec:D6onspindle} we study solutions of 8d gauged supergravity and uplift them to eleven dimensions. We start revisiting the solutions of \cite{Edelstein:2001pu}, corresponding (in 10d) to D6 branes wrapped on $S^2$ and then introduce the new spindle solutions, carefully analyzing the associated Killing spinors and pointing out the connection with $Y^{p,q}$ manifolds. In Section \ref{sec:Ypq} we generalize the story to $Y^{p,q}$ manifolds, which we view as $U(1)$ fibrations over certain K\"hler orbifolds and we exhibit the associated Killing spinors. Finally, in Section \ref{sec:resolution} we show that the metrics describing the small resolution of the Calabi-Yau cone over $Y^{p,q}$ can also be obtained from the uplift of solutions in 8d gauged supergravity. We conclude with some discussion and outlook in Section \ref{sec:discussion}, while we describe our notation for the gamma matrices in Appendix \ref{app:gammas}.

\section{D6 branes wrapped on a spindle}\label{sec:D6onspindle}

\subsection{8d gauged supergravity and its uplift}\label{sec:8dsugra}

We consider the 8d maximal $SU(2)$-gauged supergravity constructed in \cite{Salam:1984ft} and focus on the truncation discussed in \cite{Edelstein:2001pu}, which keeps only the metric, a gauge field $A$ (with $F=\diff A$) and two scalars $\varphi$ and $\phi$. The action for the bosonic fields is
\es{8daction}{
S_{\mathrm{8d}}=\frac{1}{16\pi G^{(8)}_N}\int \diff^8x\sqrt{-g}\left[R+\mathcal{V}-\frac{1}{2}(\partial\varphi)^2-6(\partial\phi)^2-\frac{1}{4}\ex^{\varphi+4\phi}(F_{\mu\nu})^2\right]\,,
}
where 
\es{V8d}{
\mathcal{V}=\frac{g^2}{2}\ex^{2\phi-\varphi}(4-\ex^{6\phi})\,,
}
where $g$ is the gauge coupling. A solution to this model preserves supersymmetry if the following Killing spinor equations 
\es{KSE}{
\delta\psi_\mu=&\left[\nabla_\mu-\frac{\ii}{2}\,g\,A\,\gammathreeE_3+\frac{1}{48}\ex^{2\phi+\varphi/2}\left(\gammaeightL_{\mu\nu\rho}-10 \delta_{\mu}^{\,\,\nu}\gammaeightL_\rho\right)\gammaeightL_\star\,F^{\nu\rho}\,\gammathreeE_3\right.\\
&\left.\,\,\,+\frac{\ii}{24}\,g\,\ex^{-\varphi/2}(\ex^{4\phi}+2\ex^{-2\phi})\gammaeightL_\mu\,\gammaeightL_\star\right]\epsilon=0\,,\\
\delta\chi=&\left[\slashed{\partial}\phi+\frac{1}{6}\slashed{\partial}\varphi-\frac{\ii}{4}\,g\,\ex^{-\varphi/2}(\ex^{4\phi}-\ex^{-2\phi})\,\gammaeightL_\star-\frac{1}{8}\ex^{2\phi+\varphi/2}\slashed{F}\,\gammaeightL_\star\,\gammathreeE_3\right]\epsilon=0\,,
}
are satisfied by non-trivial spinors $\epsilon$, where $\psi_{\mu}$ is the gravitino while $\chi$ is a matter fermion. Note that $\epsilon$ is a spinor of both the Lorentz group $SO(1,7)$ and of the R-symmetry group $SU(2)\simeq SO(3)/\Z_2$, whose gamma matrices we denote with $\gammathreeE$.

The 8d supergravity of \cite{Salam:1984ft} was originally found as a Scherk–Schwarz compactification of 11d supergravity on $S^3$. As such, its (supersymmetric) solutions uplift to (supersymmetric) solutions of 11d supergravity. The specific truncation we are interested in is such that solutions to \eqref{8daction} uplift to pure geometry in 11d, with metric given by
\es{uplift8dto11d}{
\ds_{11}=\ex^{-\varphi/3}\ds_8+\frac{1}{g^2}\ex^{2\varphi/3}\left[\ex^{-2\phi}(\diff\theta^2+\sin^2\theta\diff\alpha^2)+\ex^{4\phi}(\diff\psi-\cos\theta\diff\alpha+g\,A)^2\right]\,,
}
where $(\theta,\alpha,\psi)$ parameterize $S^3$ if $\theta\in[0,\pi]$, $\alpha\in[0,2\pi)$ and $\psi\in[0,4\pi)$. Note that this can be reduced along the $\psi$-circle to a solution of type IIA supergravity with metric, dilaton and RR one-form $C_1$, which in the string frame read
\es{uplift8dto10d}{
\ds_{10}=&\ex^{2\phi}\ds_8+\frac{1}{g^2}\ex^{\varphi}(\diff\theta^2+\sin^2\theta\diff\alpha^2)\,,\\
C_1=&-\cos\theta\diff\alpha+g\,A\,,\\
\Phi=&\frac{\varphi}{2}+2\phi\,,
}
which is the form of a IIA solution describing D6 branes. Indeed, the IIA solution describing the near-horizon limit of a stack of D6 branes can be found from the uplift of a solution to \eqref{8daction} whose only non-trivial fields are the metric and the scalar $\varphi$, given by
\es{D6NH_8d}{
\ds_8&=r\,\left[\ds(\R^{1,6})+\diff r^2\right]\,,\\
\ex^{\varphi}&=\frac{g^2\,r^3}{4}\,.
}

\subsection{Warm up: D6 branes wrapped on $S^2$}\label{sec:D6onS2}

We begin by reviewing the solution found in \cite{Edelstein:2001pu}, which describes D6 branes wrapped on a two-cycle $S^2$ inside the Eguchi-Hanson space. The authors consider an ansatz of the type
\es{S2ansatz_r}{
\ds_8&=\ex^{2f}\,\left[\ds(\R^{1,4})+\diff r^2+\ex^{2h}\,\ds(S^2)\right]\,,\\
F&=\frac{1}{g}\,\vol(S^2)\,,
}
where$f,h$ and the scalars $\varphi,\phi$ are functions of $r$. One of the results of \cite{Edelstein:2001pu} is the existence of a solution with non-trivial $r$-dependence which, once uplifted to 11d using \eqref{uplift8dto11d}, describes the resolved conifold. However, there is a simpler solution, which in 11d language corresponds to the vanishing limit of the resolution parameter, where the geometry reduces to the cone over the Sasaki-Einstein manifold $T^{1,1}$. In this limit, we have
\es{S2sol}{
\ds_8&=\frac{g\,r}{2}(2/3)^{2/3}\,\left[\ds(\R^{1,4})+\diff r^2+\frac{r^2}{6}\,(\diff\theta'^2+\sin^2\theta'\,\diff\alpha'^2)\right]\,,\\
A&=-\frac{1}{g}\cos\theta'\,\diff\alpha'\,,\quad
\ex^{\varphi}=\frac{g^3\,r^3}{18}\,,\quad
\ex^{6\phi}=\frac{2}{3}\,,
}
where the constant parameters have been chosen in such a way that using \eqref{uplift8dto11d} this uplifts to the conifold with the standard metric\footnote{Note that we have relabelled $(\theta',\alpha')\to (\theta_1,\alpha_1)$ and $(\theta,\alpha)\to(\theta_2,\alpha_2)$.}
\es{conifold}{
\ds_{11}=&\ds(\R^{1,4})+\diff r^2+r^2\,\ds(T^{1,1})\,,\\
\ds(T^{1,1})=&\frac{1}{6}\left(\diff\theta_1^2+\sin^2\theta_1^2\,\diff\alpha_1^2+\diff\theta_2^2+\sin^2\theta_2^2\,\diff\alpha_2^2\right)\\
&+\frac{1}{9}\left(\diff\psi-\cos\theta_1\,\diff\alpha_1-\cos\theta_2\,\diff\alpha_2\right)^2\,.
}
We also note that \eqref{S2sol} preserves eight Killing spinors $\epsilon$, which are subject to the two projections
\es{projectionsS2}{
\gammaeightL_{r\theta'\alpha'}\epsilon=\gammathreeE_3\epsilon\,,\quad
\gammaeightL_r\,\epsilon=\ii\,\gammaeightL_\star\,\epsilon\,.
}

\subsection{D6 branes wrapped on $\spindle$: local solution}

We would now like to consider solutions which describe D6 branes wrapping a spindle $\spindle$. We recall that a spindle is topologically a one (complex) dimensional weighted projective space and it does not admit metrics of constant curvature. The simplest metrics on spindles have $U(1)$ invariance and can be written in the form
\es{generalspindle}{
\ds(\spindle)=\diff y^2+f(y)\,\diff z^2\,,
}
where $f(y)$ vanishes at $y=y_{1,2}$ and is positive for $y\in (y_1,y_2)$. The space has quantized conical deficits $2\pi(1-n_{1,2}^{-1})$ at the two poles $y=y_{1,2}$, characterized by two coprime integers $n_{1,2}\in \N$. See \cite{Ferrero:2021etw} for more details.

Naively, one would think that the natural way to proceed would be to generalize the ansatz \eqref{S2ansatz_r} keeping an arbitrary dependence on $r$ and adding a non-trivial dependence on $y$ on all functions. In that case, the Killing spinor equations would lead to a system of non-linear partial differential equations: a highly complicated problem, which could only be solved with some inspired guesswork.\footnote{As we will discuss in Section \ref{sec:resolution}.} Instead, we consider the following approach: we consider the simpler solution \eqref{S2sol} as a starting point for an ansatz, keeping the same dependence on $r$ for all functions but allowing for an arbitrary dependence on $y$. See \cite{Boisvert:2024jrl} for additional comments on this approach and other examples.

After some trial and error, we land on the ansatz\footnote{Note that the choice of gauge is such that the Killing spinors are independent of $z$.}
\es{8dspindle}{
\ds_8&=r\,(y\,H)^{1/6}\,\left[\ds(\R^{1,4})+\diff r^2+r^2\,\ds(\spindle)\right]\,,\quad
\ds(\spindle)=\frac{1}{g^2}\frac{y}{P}\diff y^2+\frac{g^2}{4}\frac{P}{H}\diff z^2\,,\\
A&=\left(\frac{q}{H}+\frac{1}{2}\right)\,\diff z\,,\quad
\ex^{\varphi}=(y\,H)^{1/2}\,r^3\,,\quad
\ex^{6\phi}=\frac{H}{y^2}\,,
}
where the functions are
\es{8dfunctions}{
H=y^2+q\,,\quad 
P=H-\frac{4}{g^2}\,y^3\,.
}
We note that this solution reduces to \eqref{S2sol} in a scaling limit, in the spirit of Appendix A of \cite{Ferrero:2021etw}. To show this, we set
\es{scalingS2}{
y\to \frac{g^2}{6}+\delta\,\cos\theta'\,,\quad
z\to-\frac{g}{9}\frac{\alpha'}{\delta}\,,\quad
q\to-\frac{g^4}{108}+\delta^2\,,
}
and take the limit $\delta\to 0$. Note that $P\sim O(\delta^2)$ in this limit.

\subsection{Killing spinors}

We will now show explicitly that the solution \eqref{8dspindle} preserves eight Killing spinors, which we exhibit explicitly. To do so, we introduce the frame
\es{frame}{
\ex^{\mu}=\Omega\, \ext^\mu\,,\quad
\Omega=\sqrt{r}\,(y\,H)^{1/12}\,,\quad
\mu=0,\ldots,7\,,
}
where
\es{frametilde}{
\ext^{\alpha}=\diff x^{\alpha}\,,\quad (\alpha=0,\ldots,4)\,,\quad
\ext^r=\diff r\,,\quad
\ext^y=\frac{1}{g}\frac{\sqrt{y}}{\sqrt{P}}\,\diff y\,,\quad
\ext^z=\frac{g}{2}\frac{\sqrt{P}}{\sqrt{H}}\diff z\,,
}
and $x^{\alpha}$ ($\alpha=0,\ldots,4$) are Cartesian coordinates on $\R^{1,4}$. With this choice of frame, the eight Killing spinors $\epsilon$ preserved by the solution are subject to the projections
\es{projectionsspindle}{
\gammaeightL_{ryz}\epsilon=\gammathreeE_3\epsilon\,,\quad
\left(\sin\beta\,\gammaeightL_y+\cos\beta\,\gammaeightL_r\right)\,\epsilon=\ii\,\gammaeightL_\star\,\epsilon\,,
}
where we have introduced the angle $\beta$ via
\es{betadef}{
\sin\beta=\frac{\sqrt{P}}{\sqrt{H}}\,,\quad
\cos\beta=\frac{2y/g}{\sqrt{H}}\,.
}
Note that the first of the two projections in \eqref{projectionsspindle} is the same as the first in \eqref{projectionsS2}, while the second reduces to \eqref{projectionsS2} when $\beta=0$, which can be obtained for $P=0$. This is precisely what is achieved by the scaling limit \eqref{scalingS2}, since $P\sim O(\delta^2)$ in that limit.

To give the explicit expression of the spinors, it is convenient to use the basis of 8d spacetime gamma matrices $\gammaeightL_\mu$ and 3d flavor gamma matrices $\gammathreeE_i$ (see Appendix \ref{app:gammas} for our conventions). The necessary ingredients are 6d spinors $\chi_\pm$ satisfying
\es{chi6d}{
\gammasixL_r\,\chi_\pm=\pm \ii\,\gammasixL_\star \chi_\pm\,,
}
and 3d flavor spinors $\xi_\pm$ satisfying
\es{xi2d}{
\gammathreeE_3\,\xi_\pm=\pm\,\xi_\pm\,.
}
Finally, we define the spinor $\zeta$ on $\spindle$ and its conjugate $\zeta^c$ as
\es{zeta}{
\zeta=
\begin{pmatrix}
\cos\frac{\beta}{2}\\
\ii\,\sin\frac{\beta}{2}
\end{pmatrix}\,,
\quad
\zeta^c=B\,\zeta^*=
\begin{pmatrix}
-\ii\,\sin\frac{\beta}{2}\\
\cos\frac{\beta}{2}
\end{pmatrix}\,,
}
where the charge conjugation matrix $B=\sigma^1$ is such that $B\,(\sigma^i)^*\,B^{-1}=\sigma^i$ ($i=1,2,3$). We note that the components of $\zeta$ are conveniently expressed in terms of two functions $h_\pm$ 
\es{hpm}{
h_\pm=\sqrt{H}\pm\frac{2\,y}{g}\,,
}
via
\es{betahalf}{
\sin\frac{\beta}{2}=\frac{\sqrt{h_-}}{\sqrt{h_++h_-}}\,,\quad
\cos\frac{\beta}{2}=\frac{\sqrt{h_+}}{\sqrt{h_++h_-}}\,.
}
In terms of these building blocks, the eight Killing spinors $\epsilon$ preserved by the solution are given by
\es{8dspinors}{
\epsilon=\sqrt{\Omega}\,\left[\chi_+\otimes \zeta\otimes \xi_++\chi_-\otimes \zeta^c\otimes \xi_-\right]\,.
}
We conclude by observing that in the case $q=0$ the number of Killing spinors doubles. Their expression is still of the form \eqref{8dspinors}, but with $\chi_\pm$ replaces with two independent arbitrary 6d Dirac spinors.

\subsection{Global analysis}\label{sec:globalspindle}

Let us now consider the conditions on the range of the coordinates and on the parameter $q$ such that the space parametrized by $y, z$ parametrizes a spindle $\spindle$. We note that $P$ is a polynomial of degree three where the coefficient of the highest degree term is negative. Thus, the only way to have $P>0$ in a compact region is that all three roots of $P$ are real. Using Descartes' rule of signs, it is immediate to show that this is possible only for $q<0$, in which case there is one negative root ($y_0<0$) and either zero or two positive roots ($0<y_1<y_2$). It is also straightforward to show that $y_{1,2}$ are real only for $q>-g^4/108$, so we end up with the conditions
\es{qconstr}{
-\frac{g^4}{108}<q<0\,, \quad y_0<0<y_1<y_2\,.
}
We can study the behavior of the metric $\ds(\spindle)$ near the roots $y_i$ by setting
\es{ytorho}{
y\to y_i+\frac{g^2\,P'(y_i)}{4y_0}\,\rho^2\,,
}
and expanding for small $\rho$. We find
\es{spindlemetricexp}{
\ds(\spindle)\simeq \diff\rho^2+k_i^2\,\rho^2\,\diff z^2\,,\quad 
k_i=\left|\frac{g^2\,P'(y_i)}{4\sqrt{y_i\,H(y_i)}}\right|\,.
}
We can simplify the expression of $k_i$ observing that $y_i>0$ and $H(y_i)=\frac{4}{g^2}y_i^3$, moreover $P'(y_1)>0$ and $P'(y_2)<0$ which gives
\es{kappai}{
k_1=\frac{g^3\,P'(y_1)}{8y_1^2}\,,\quad
k_2=-\frac{g^3\,P'(y_2)}{8y_2^2}\,.
}
We require that $\spindle$ has quantized conical deficits at $y=y_{1,2}$, which is achieved when the periodicity $\Delta z$ of $z$ is
\es{Deltaz12}{
\Delta z=\frac{2\pi}{k_1\,n_1}=\frac{2\pi}{k_2\,n_2}\,.
}
It is also straightforward to show that
\es{A12}{
\left.A\right|_{y=y_i}=-\frac{g^2\,P'(y_i)}{8y^2}\,,
}
from which we find that the flux of $A$ across the spindle evaluates to
\es{Qflux}{
Q=\frac{g}{2\pi}\int_\spindle\diff A=\frac{\Delta z}{2\pi}(k_1+k_2)=\frac{n_1+n_2}{n_1\,n_2}\,,
}
showing that only twist solutions are realized in this class \cite{Ferrero:2021etw}. The conditions \eqref{Deltaz12} can be solved explicitly to find
\es{spindlepar8d}{
q&=-\frac{g^4}{216}\left(1-\frac{(n_1^2-4n_1n_2+n_2^2)}{(n_1+n_2)^3}\sqrt{n_1^2+14n_1n_2+n_2^2}\right)\,,\\
\Delta z&=\frac{2\pi}{9g\,n_1n_2(n_1-n_2)}\left[(n_1^2-10n_1n_2+n_2^2)-(n_1+n_2)\sqrt{n_1^2+14n_1n_2+n_2^2}\right]\,,
}
while the roots are given by
\es{spindleroots8d}{
y_1&=\frac{g^2}{4}\frac{n_1}{n_1+n_2}-\frac{1}{2}y_0\,,\quad
y_2=\frac{g^2}{4}\frac{n_2}{n_1+n_2}-\frac{1}{2}y_0\,,\\
y_0&=\frac{g^2}{12}\left(1-\frac{\sqrt{n_1^2+14n_1n_2+n_2^2}}{n_1+n_2}\right)\,.
}
Note that the ordering $y_0<y_1<y_2$ is respected whenever $n_2>n_1$, and moreover we always have $-g^4/108<q<0$.

We conclude with a comment on the case $q=0$, which gives a doubling of supersymmetry and resembles the disk solutions that have appeared in the literature starting with \cite{Bah:2021mzw}. A crucial difference is that the gauge field flux vanishes in this case. For the metric functions we find
\es{functiondisk}{
H=y^2\,,\quad
P=\frac{y^2}{g^2}(g^2-4y)\,,
}
which gives a compact range with positive definite signature only taking
\es{rangedisk}{
0<y<\frac{g^2}{4}\,.
}
At $y=g^2/4$ the $z$-circle shrinks and one can choose whether to have a conical deficit or to end the space smoothly, while at $y=0$ the circle does {\it not} shrink and the space ends with a boundary where there is a curvature singularity. As we shall soon see, this singularity is only present in gauged supergravity and is completely cured by the uplift to 11d.

\subsection{Uplift to 11d}

Let us consider the uplift of the solution \eqref{8dspindle} to 11d. The metric reads
\es{uplift8dspindle}{
\ds_{11}=&\ds(\R^{1,4})+\diff\rr^2+\frac{\rr^2}{g^2}\ds(Y_5)\,,\\
\ds(Y_5)=&\frac{y}{P}\diff y^2+\frac{g^4\,P}{4\,H}\diff z^2+y\,(\diff\theta^2+\sin^2\theta\diff\alpha^2)+\frac{H}{y}(\diff\psi^2-\cos\theta\,\diff\alpha+g\,\frac{\q}{H}\,\diff z)^2\,,
}
where supersymmetry and the equations of motion require $Y_5$ to be a Sasaki-Einstein manifold. As it turns out, $\ds(Y_5)$ is diffeomorphic to the $Y^{p,q}$ metrics found in \cite{Gauntlett:2004yd,Gauntlett:2004hh}, as it can be shown with the change of coordinates
\es{comparetoYpq}{
y\to \frac{1-c\,y'}{6}g^2\,,\quad
\psi\to-\frac{\beta-2c\,\psi'}{3c}\,,\quad
z\to -\frac{2(\beta+c\,\psi')}{3c\,g}\,,
}
which leads to the same metric given in eq. (4.1) of \cite{Gauntlett:2004yd}.\footnote{The coordinates $y'$ and $\psi'$ here are $y$ and $\psi$ of \cite{Gauntlett:2004yd}.} We thus obtain a new perspective on five-dimensional $Y^{p,q}$ Sasaki-Einstein manifolds, which here are viewed as a circle fibration over a base space which is the warped product of $\mathbb{CP}^1$ and $\spindle$, where the circle fibration realizes the canonical bundle over $\mathbb{CP}^1$ and the twist over $\spindle$. We will consider this perspective for $Y^{p,q}$ manifolds of general dimension in the next section.

We conclude with a comment on the $q=0$ case. The change of coordinates
\es{diskS5}{
y\to \frac{g^2}{4}\sin^2\sigma\,,\quad
z\to \frac{2}{3g}\xi\,,\quad
\psi\to\frac{2}{3}\xi-\beta\,,
}
shows that $\ds(Y_5)$ in \eqref{uplift8dspindle} with $\q=0$ becomes
\es{S5metric}{
\ds(S^5)=&\diff\sigma^2+\frac{1}{4}\sin^2\sigma\,(\diff\theta^2+\sin^2\theta\,\diff\alpha^2)+\frac{1}{4}\sin^2\sigma\,\cos^2\sigma\,(\diff\beta+\cos\theta\,\diff\alpha)^2\\
&+\frac{1}{9}\left(\diff\xi-\frac{3}{2}\sin^2\sigma\,(\diff\beta+\cos\theta\,\diff\alpha)\right)^2\,,
}
that is the round metric on a five sphere with unit radius, if we choose $\sigma\in [0,\tfrac{\pi}{2})$, $\theta\in[0,\pi)$, $\alpha \sim \alpha+2\pi$, $\beta\sim \beta+4\pi$ and $\xi\sim \xi+6\pi$. \footnote{Here we are viewing $S^5$ as a $U(1)$ fibration over $\mathbb{CP}^2$, equipped with the Fubini-Study metric, where for the homogeneous coordinates $\zeta_1=\frac{z_1}{z_3}$, $\zeta_2=\frac{z_2}{z_3}$ describing the embedding of $\mathbb{CP}^2$ into $\C^3$ parametrized by $z_{1,2,3}$, we set
\es{}{
\zeta_1=\tan\sigma\,\cos\tfrac{\theta}{2}\,\ex^{\ii\frac{\alpha+\beta}{2}}\,,\quad
\zeta_2=\tan\sigma\,\sin\tfrac{\theta}{2}\,\ex^{\ii\frac{\alpha-\beta}{2}}\,.
} This is compatible with $|z_3|=\cos\sigma$ and the additional angle $\xi$ is defined by $z_3=\cos\sigma\,\ex^{\ii\frac{\xi}{3}}$.} The curvature singularity present in 8d gauged supergravity is therefore completely absent after the uplift.

\section{$Y^{p,q}$ manifolds from spindles}\label{sec:Ypq}

We now show that a similar perspective generalizes to $Y^{p,q}$ Sasaki-Einstein manifolds of general dimension. We start by writing their metrics as a fibration over a four-dimensional orbifold which is the warped product of a spindle and a K\"ahler-Einstein manifold and we exhibit the associated Killing spinors explicitly. We then conclude discussing how this perspective can be used to obtain other new solutions of 8d gauged supergravity.

\subsection{Rethinking $Y^{p,q}$ metrics}

Let us begin by reviewing $Y^{p,q}$ metrics in general dimensions, as they were originally presented in \cite{Gauntlett:2004hh}. The starting point is the local metric in $2n+2$ dimensions
\es{KEmetric}{
\diff\hat{s}^2=U^{-1}\,\diff\rho^2+\rho^2\,U(\diff\tau-\tilde{\sigma})^2+\rho^2\,\diff\tilde{s}^2\,,
}
where $\diff\tilde{s}^2$ is the metric on a $2n$-dimensional K\"ahler-Einstein manifold $B_{2n}$ with $\widetilde{\text{Ric}}=\lambda\,\tilde{g}$, $\lambda>0$ and K\"ahler form $\tilde{J}=\tfrac{1}{2}\diff \tilde{\sigma}$. When the function $U$ is given by
\es{Urho}{
U=\frac{\lambda}{2(n+1)}-\frac{\Lambda}{2(n+2)}\rho^2+\frac{\Lambda}{2(n+1)(n+2)}\left(\frac{\lambda}{\Lambda}\right)^{n+2}\frac{\kappa}{\rho^{2n+2}}\,,
}
$\diff\hat{s}^2$ in \eqref{KEmetric} describes a positive curvature K\"ahler-Einstein metric with $\widehat{\text{Ric}}=\Lambda\,\hat{g}$, $\Lambda>0$, and K\"ahler form $\hat{J}$ given by
\es{Jhat}{
\hat{J}=\rho^2\,\tilde{J}+\rho\,(\diff\tau-\tilde{\sigma})\wedge\diff\rho\,.
}
The local form of $Y^{p,q}$ metrics is then given by
\es{lineYpq}{
\ds(Y^{p,q})=(\diff\xi+\hat{\sigma})^2+\diff\hat{s}^2\,,
}
which is locally Sasaki-Einstein with curvature $2(n+1)$ provided $\Lambda=2(n+2)$. Here $\diff\hat{\sigma}=2\hat{J}$ and a convenient choice is
\es{sigma}{
\sigma=\frac{\lambda}{\Lambda}\tilde{\sigma}+\left(\frac{\lambda}{\Lambda}-\rho^2\right)(\diff\tau-\tilde{\sigma})\,.
}
With the simple change of coordinates
\es{anglesYpq}{
\xi\to z-\frac{\lambda}{\Lambda}\,\psi\,,\quad \tau \to \psi\,,
}
we can rewrite the metric \eqref{lineYpq} as
\es{lineYpqnew}{
\ds(Y^{p,q})=\rho^2\,(U+\rho^2)\,\left(\diff\psi-\tilde{\sigma}-A\right)^2+\frac{\diff\rho^2}{U}+\frac{U}{(U+\rho^2)}\diff z^2+\rho^2\,\diff\tilde{s}^2\,,\quad A=\frac{\diff z}{(U+\rho^2)}\,.
}
The metric \eqref{lineYpqnew} is written as a circle fibration over the warped product between $B_{2n}$ and a spindle $\sigma$ parametrized by $\rho$ and $z$. We find it useful to consider a further change of coordinates, inspired by \cite{Gauntlett:2004hh}. We set
\es{rhotoy}{
\rho \to \sqrt{\frac{\lambda}{\Lambda}}\sqrt{y}\,,
}
and shift $\psi \to \psi+\tfrac{\Lambda}{\lambda} z$, to obtain the final form of the $Y^{p,q}$ metric as\footnote{The effect of the shift in $\psi$ is that of fixing a gauge for the connection $A$, which is such that the Killing spinors are independent of $z$.}
\es{lineYpqfinal}{
\ds(Y^{p,q})&=\frac{\lambda^2}{\Lambda^2}\frac{H}{(n+1)y^n}\left(\diff\psi-\tilde{\sigma}-A\right)^2+\frac{\lambda}{\Lambda}\,y\,\diff\tilde{s}^2+\ds(\spindle)\,,\\
\ds(\spindle)&=\frac{n+1}{4}\frac{y^n}{P}\diff y^2+\frac{P}{H}\diff z^2\,,\\
A&=\frac{2}{\lambda}\,\left((n+1)\frac{\kappa}{H}+1\right)\,,
}
where we have used $\Lambda=2(n+2)$ and the functions $P$ and $H$ are given by
\es{PHYpq}{
H=(n+2)\,y^{n+1}+\kappa\,,\quad
P=H-(n+1)\,y^{n+2}\,.
}
Note that the notation here and in the next subsections is intentionally very similar to that of Section \ref{sec:8dsugra}, in order to ease the comparison between the formulas.

We would now like to consider the conditions for global regularity of \eqref{lineYpqfinal}, from a spindle perspective. First we focus on the fibration over $B_{2n}$: since $\diff\tilde{\sigma}=2\tilde{J}$ and $\widetilde{\text{Ric}}=\lambda\,\tilde{g}$, we obtain the $U(1)$ fibration associated with the canonical bundle over $B_{2n}$ if the periodicity of $\psi$ satisfies
\es{Deltapsi}{
\Delta\psi=\frac{4\pi}{\lambda}\,.
}
Then we would like to obtain a compact range of $y$ where $P>0$ and $H>0$. Moreover, the warping factor $y$ of the $B_{2n}$ metric $\diff\tilde{s}^2$ requires $y>0$.\footnote{This condition can be dropped for $n=0$, which we discuss in Section \ref{sec:dim3Ypq}.} The same analysis performed in Section 2 of \cite{Gauntlett:2004hh} shows that the only way to satisfy all conditions is to take
\es{kapparange}{
-1<\kappa\le 0\,,
}
in which case $P$ has at least two real roots $y_1$ and $y_2$ satisfying
\es{yroots}{
0\le y_1<1<y_2\le \sqrt{\frac{n+2}{n+1}}\,.
}
We then focus on this case and consider the expansion of the spindle metric near the roots $y_{1,2}$. We set $y\to y_i+(n+2)(1-y_i)s^2$ and expanding for small $s$ we obtain
\es{}{
\ds(\spindle)\simeq \diff s^2+k_i^2\,s^2\,\diff z^2\,,
}
where the condition \eqref{yroots} fixes the sign of $k_i$ and we have
\es{kappa12}{
k_1=(n+2)\frac{1-y_1}{y_1}\,,\quad
k_2=(n+2)\frac{y_2-1}{y_2}\,.
}
We obtain a smooth metric on $\spindle$ with quantized conical deficits at the poles if
\es{Deltaz}{
\Delta z=\frac{2\pi}{k_1\,n_1}=\frac{2\pi}{k_2\,n_2}\,.
}
This allows to compute the Euler number of the spindle directly in the usual way, without explicitly solving \eqref{Deltaz}. We find
\es{euler}{
\chi(\spindle)=\frac{1}{2\pi}\int_{\spindle}\diff\omega_{\spindle}=\frac{n_1+n_2}{n_1\,n_2}\,,
}
where $\omega_{\spindle}$ is the spin connection associated with $\ds(\spindle)$. The last condition that we should examine is the regularity of the $U(1)_\psi$ orbi-bundle over $\spindle$, with connection one-form $A$. We find that using $H(y_i)=(n+1)y_i^{n+2}$ we can write
\es{ApolesYpq}{
\left.A\right|_{y=y_i}=\frac{2}{\lambda}\frac{(n+2)(y_i-1)}{y_i}\diff z\,,
}
which using \eqref{Deltaz} allows to compute
\es{fluxAYpq}{
\frac{1}{\Delta \psi}\int_{\spindle}\diff A=\frac{n_1+n_2}{n_1\,n_2}\,,
}
which shows that $A$ is a connection on $\mathcal{O}(n_1+n_2)$, {\it i.e.} the tangent bundle over $\spindle$. This was called the twist case in \cite{Ferrero:2021etw}.

The case of three-dimensional Sasaki-Einstein manifolds is special and essentially gives metrics on 3d Lens spaces. We note that the requirement \eqref{kapparange} on the range of $\kappa$ comes from the condition that $y>0$ in \eqref{lineYpqfinal}, which is in turn dictated by the warping factor $y$ in front of $\diff\tilde{s}^2$. However, that part of the metric is not present in the 3d case ($n=0$), in which the metric reads (after setting $\lambda=2$ for convenience)
\es{ds3}{
\ds_3&=\ds(\spindle)+\frac{H}{4}\,(\diff \psi+A)^2\,,\quad 
A=\left(\frac{\kappa}{H}+1\right)\,\diff z\,,\\
\ds(\spindle)&=\frac{\diff y^2}{4P}+\frac{P}{H}\diff z^2\,,
}
with 
\es{}{
H=2y+\kappa\,,\quad P=H-y^2\,.
}
Now $P$ has exactly two roots $y_{1,2}$, whose expression is
\es{}{
y_1=1-\sqrt{1+\kappa}\,,\quad
y_2=1+\sqrt{1+\kappa}\,.
}
Note that as in the previous subsection we have $0<y_1<1<y_2<2$ for $-1<\kappa<0$, which realizes the twist as previously discussed. The conditions for global regularity can be solved explicitly and give
\es{}{
\kappa=-\frac{4n_1n_2}{(n_1+n_2)^2}\,,\quad
\Delta z=\frac{2\pi}{n_2-n_1}\,,
}
as well as
\es{}{
y_1=\frac{2n_1}{n_1+n_2}\,,\quad
y_2=\frac{2n_2}{n_1+n_2}\,,
}
for the roots, and note that $y_1<y_2$ requires $n_1<n_2$. In this case, as anticipated, the flux of $A$ across the spindle is
\es{}{
\frac{1}{2\pi}\int \diff A=\frac{n_1+n_2}{n_1\,n_2}\,,
}
which is the twist condition.

However, now there is no obstruction in considering the situation
\es{}{
\kappa>0,\quad
y_1<0<y_2\,.
}
The conditions for the quantization of the conical deficits are easily solved by
\es{}{
\kappa=\frac{4n_1n_2}{(n_2-n_1)^2}\,,\quad
\Delta z=\frac{2\pi}{n_1+n_2}\,,
}
which allows to rewrite the roots as
\es{}{
y_1=-\frac{2n_1}{n_2-n_1}\,,\quad
y_2=\frac{2n_2}{n_2-n_1}\,,
}
which satisfy the ordering $y_1<y_2$ for $n_1<n_2$. The point that we want to make here is that the flux of the connection $A$ in this case is
\es{}{
\frac{1}{2\pi}\int \diff A=\frac{n_1-n_2}{n_1\,n_2}\,,
}
that is the fibration realizes the anti-twist condition \cite{Ferrero:2021etw}. We would also like to emphasize that the metric \eqref{ds3} can {\it locally} be obtained from the round metric on $S^3$: indeed, with the change of coordinates
\es{}{
y\to\sqrt{1+\kappa}\cos\theta+1\,,\quad
z\to -\frac{1}{2}\left(\frac{\alpha}{\sqrt{1+\kappa}}+\psi\right)\,,
}
one obtains
\es{}{
\ds_3=\frac{1}{4}\left[\diff\theta^2+\sin^2\theta\,\diff\alpha^2+(\diff\psi-\cos\theta\,\diff\alpha)\right]\,.
}
However, this only applies locally, while the global structure of the 3d manifold \eqref{ds3} is generally that of a lens space. A carefuly global analysis of this space was carried out in Section 5 of \cite{Inglese:2023tyc}, where the connection between spindles and 3d lens spaces is used to study supersymmetric partition functions on the spindle. We refer the reader to that paper for a detailed analysis of the relation between spindles and (squashed) 3d lens spaces.

\subsection{Killing spinors}

We now consider the Killing spinor equation associated with Sasaki-Einstein manifolds. We show that if we consider a frame adapted to the metric \eqref{lineYpqfinal}, it is possible to express the associated solutions as a tensor product of spinors on the K\"ahler-Einstein manifold $B_{2n}$ and spinors on the spindle. This shows the non-trivial fact that reducing on $\partial_{\psi}$ in \eqref{lineYpqfinal} gives rise to a {\it supersymmetric} K\"ahler orbifold. The expression for the spinors on the spindle parallels the one that can be found in the numerous spindle solutions in gauged supergravity that have appeared in the literature.

We begin by introducing the frame
\es{frameYpq}{
\ex^i&=y\,\sqrt{\frac{\lambda}{\Lambda}}\,\ex^i_{B_{2n}}\,,\quad (i=0,\ldots,2n)\,,\quad
\ex^y=\frac{\sqrt{n+1}}{2}\frac{y^{n/2}}{\sqrt{P}}\diff y\,,\quad
\ex^z=\frac{\sqrt{P}}{\sqrt{H}}\diff z\,,\\
\ex^\psi&=\frac{\lambda}{\Lambda}\frac{\sqrt{H}}{\sqrt{n+1}y^{n/2}}\left(\diff\psi-\tilde{\sigma}-A\right)\,.
}
A convenient choice of gamma matrices $\tau^{(2n+3)}_a$, $a=1,\ldots,2n+3$, is then
\es{gammasgeneral}{
\tau^{(2n+3)}_i&=\tau^{(2n)}_i\otimes 1\,,\quad
(i=1,\ldots,2n)\,,\\
\tau^{(2n+3)}_y&=\tau^{(2n)}_\star\otimes \sigma^1\,,\quad
\tau^{(2n+3)}_z=\tau^{(2n)}_\star\otimes \sigma^2\,,\quad
\tau^{(2n+3)}_\psi=\tau^{(2n)}_\star\otimes \sigma^3\,,
}
where $\tau^{(2n)}_i$ are gamma matrices in $2n$ dimensions and $\tau^{(2n)}_\star$ the associated chirality matrix.

Every complete $(2n+3)$-dimensional Sasaki-Einstein manifold admits at least two linearly independent Killing spinors $\epsilon$, which satisfy
\es{KSESE}{
\nabla_a\,\epsilon=\ii\,\frac{\eta}{2}\,\tau^{(2n+3)}_a\,\epsilon\,,
}
where $\eta=1$ for both spinors if $n=2p$ while $\eta=1,-1$ for the two spinors if $n=2p+1$, $p\in \N$. The two solutions of \eqref{KSESE} can be expressed in the general case as
\es{spinorsYpq}{
\epsilon_+=\ex^{\ii\frac{\psi'}{2}}\chi_+\otimes \zeta\,,\quad
\epsilon_-=\ex^{-\ii\frac{\psi'}{2}}\chi_-\otimes \zeta^c\,,
}
where $\chi_\pm$ are positive and negative chirality Killing spinors on the K\"ahler-Einstein manifold $B_{2n}$, $\psi'$ is a $2\pi$-periodic coordinate related to $\psi$ via
\es{}{
\psi'=\frac{2\pi}{\Delta\psi}\psi=\frac{\lambda}{2}\psi\,,
}
while $\zeta$ and its charge conjugate $\zeta^c=\sigma^1(\zeta^*)$ are Killing spinors on $\spindle$. Their explicit expression is
\es{zetaYpq}{
\zeta=
\begin{pmatrix}
\cos\frac{\beta}{2}\\
-\ii\,\sin\frac{\beta}{2}
\end{pmatrix}\,,\quad
\zeta^c=
\begin{pmatrix}
\ii\,\sin\frac{\beta}{2}\\
\cos\frac{\beta}{2}
\end{pmatrix}\,,
}
where the angle $\beta$ is the analogue of that in Section \ref{sec:8dsugra} and is defined by
\es{betahere}{
\sin\frac{\beta}{2}=\frac{\sqrt{h_-}}{\sqrt{h_++h_-}}\,,\quad
\cos\frac{\beta}{2}=\frac{\sqrt{h_+}}{\sqrt{h_++h_-}}\,,
}
where
\es{hpmhere}{
h_\pm=\sqrt{H}\pm\sqrt{n+1}\,y^{1+n/2}\,.
}
We also have the projection
\es{}{
\mathcal{M}\epsilon=\epsilon\,,\quad
\mathcal{M}=\sin\beta\,\tau^{(2n+3)}_z+\cos\beta\,\tau^{(2n+3)}_\psi\,,
}
where from \eqref{betahere} and \eqref{hpmhere} we have
\es{}{
\sin \beta=\frac{\sqrt{P}}{\sqrt{H}}\,,\quad
\cos\beta=\frac{\sqrt{n+1}y^{1+n/2}}{\sqrt{H}}\,.
}

\subsection{Back to 8d supergravity}

We conclude this section pointing out that the new perspective on $Y_{p,q}$ manifolds described in this section automatically gives new solutions of 8d gauged supergravity describing wrapped D6 branes, by simply reversing the logic of Section \ref{sec:8dsugra}. There, we found $Y^{p,q}$ manifolds arising from the 11d uplift of certain 8d solutions which in type IIA describe D6 branes wrapped on a spindle. Now we can start from 11d supergravity on the cone over a $Y^{p,q}$ manifold of dimension $2n+3$ written in the form \eqref{lineYpqfinal}, where we take
\es{CP1KE}{
\diff\tilde{s}^2=\ds(\mathbb{CP}^1)+\ds(\text{KE}_{2n-2})\,.
}
The reduction on the circle parametrized by $\psi$ in \eqref{lineYpqfinal} gives the IIA solution from the 11d one, while the further reduction on the $\mathbb{CP}^1$ appearing in \eqref{CP1KE} finally gives a solution of 8d supergravity. This describes D6 branes wrapped on the (warped) product between a spindle and a $(2n-2)$-dimensional K\"ahler-Einstein manifold, while the $\mathbb{CP}^1$ in \eqref{lineYpqfinal} plays the role of the sphere tranverse to the D6 branes in the IIA picture. As a concrete example, we can consider the 11d solution $\R^{1,2}\times C(\text{SE}_7)$ where $C(\text{SE}_7)$ is the Calabi-Yau cone over a $Y^{p,q}$ manifold of dimension 7, whose metric we write in the form \eqref{lineYpqnew} with $n=2$, and we pick $\diff\tilde{s}^2$ to be the K\"ahler-Einstein metric on $\mathbb{CP}^1\times \mathbb{CP}^1$. Then, we can reduce from 11d to IIA along the direction $\psi$ in \eqref{lineYpqnew} and then further reduce from IIA to 8d along one of the $\mathbb{CP}^1$, which is thought of as the $S^2$ in the uplift/reduction formulas \eqref{uplift8dto10d}. The result is a solution of the 8d gauged supergravity model presented in Section \ref{sec:8dsugra} where the fields read
\es{}{
\ds_8&=\frac{2^{2/3}}{3^{2/3}}\,r\,H^{1/6}\,\left[\ds(\R^{1,2})+\diff r^2+r^2 \left(\frac{y^2\,\diff y^2}{g^2\,P}+\frac{P}{H}\diff z^2+\frac{2y}{3g^2}\,(\diff \theta'^2+\sin^2\theta'\,\diff\alpha'^2)\right)\right]\,,\\
A&=\frac{1}{g}\left[\cos\theta'\,\diff\alpha'+3 \frac{q}{H}\diff z+\diff z\right]\,,\quad
\ex^{2\varphi}=\frac{2^4}{3^4}r^6\,H\,,\quad
\ex^{6\phi}=\frac{2H}{3y^3}\,,
}
and the functions are given by
\es{}{
H=y^3+q\,,\quad
P=H-\frac{4y^4}{g^2}\,.
}
It is interesting to compare this to other solutions describing branes wrapped on the product between spindles and Riemann surfaces that have appeared in the literature \cite{Cheung:2022ilc,Bomans:2023ouw,Couzens:2022lvg}. In those cases, the spindle is always fibered over the Riemann surface, which is different from the solutions obtained with the procedure described in this subsection, since the final result is a warped product (with no fibrations) between a spindle and a K\"ahler-Einstein manifold. 

We think that this comment is interesting for the following reason. In this paper we study D6 branes wrapped on two-dimensional orbifolds, which allows us to recover known metrics on Sasaki-Einstein manifolds from a new perspective. It would be interesting to extend this construction to D6 branes wrapped on more general orbifolds, such as the four-dimensional orbifolds given by spindle fibrations over other spindles ($\spindle_1\ltimes \spindle_2$) studied in \cite{Cheung:2022ilc,Bomans:2023ouw,Couzens:2022lvg}, or the more recently introduced quadrilaterals \cite{Faedo:2024upq}. If such solutions exist in the truncation of 11d supergravity that only contains field sourcing the 11d metric (and not the 11d four-form), then their uplift would automatically give potentially new Sasaki-Einstein metrics. So far we have been unable to find $\spindle_1\ltimes \spindle_2$ solutions in 8d supergravity which uplift to pure geometry in 11d, and we think that this might be related to the absence of a fibration structure in the $\spindle \times \sigmag$ solutions that we describe at the beginning of this subsection. On the other hand, during the submission of this work some very interesting solutions have appeared in \cite{Macpherson:2024frt}, see in particular appendix C. In that work, an orbifold version of $\CP^2$ is built using ideas that are analogous to the ones explored here.

\section{Small resolutions of conical CY metrics}\label{sec:resolution}

In Section \ref{sec:D6onS2} we have considered an ansatz for D6 branes wrapped on $S^2$, which was originally introduced in \cite{Edelstein:2001pu}, and we froze the $r$ dependence to some extremely simple form. This allows to reproduce the $T^{1,1}$ metric once the solution is uplifted to 11d, and we used this an inspiration to find solutions describing D6 branes wrapped on spindles, which uplift to 11d supergravity on the cone over five-dimensional $Y^{p,q}$ manifolds. However, this is just the simplest type of solutions found in \cite{Edelstein:2001pu}: indeed,  a more general dependence on $r$ in \eqref{S2ansatz_r} still gives BPS equations that can be explicitly solved. The resulting solution, as described in \cite{Edelstein:2001pu}, can be expressed as (after a change of coordinates)
\es{8dresolvedconifold}{
\ds_8&=\frac{g\,r}{2}(2/3)^{2/3}\,\left[\ds(\R^{1,4})+\frac{\diff r^2}{F(r)}+\frac{r^2+6a^2}{6}(\diff\theta'^2+\sin^2\theta'\diff\alpha'^2)\right]\,,\\
A&=-\frac{1}{g}\cos\theta'\diff\alpha'\,,\quad
\ex^{\varphi}=\frac{g^3\,r^3}{18}\sqrt{F(r)}\,,\quad
\ex^{\phi}=\frac{2}{3}F(r)\,,
}
where the function $F(r)$ is
\es{}{
F(r)=\frac{r^2+9a^2}{r^2+6a^2}\,,
}
and we note that this solution reduces to \eqref{S2sol} in the limit $a\to 0$, which we can also think as a limit of large $r$. What is interesting about \eqref{8dresolvedconifold} is that its uplift to 11d gives the metric
\es{resolvedconifold}{
\ds_{11}=&\ds(\R^{1,4})+\frac{\diff r^2}{F(r)}+\frac{r^2+6a^2}{6}(\diff\theta_1^2+\sin^2\theta_1^2\,\diff\alpha_1^2)+\frac{r^2}{6}(\diff\theta_2^2+\sin^2\theta_2^2\,\diff\alpha_2^2)\\
&+\frac{r^2\,F(r)}{9}\left(\diff\psi-\cos\theta_1\,\diff\alpha_1-\cos\theta_2\,\diff\alpha_2\right)^2\,,
}
which is the small resolution of the conifold $C(T^{1,1})$, where the two-sphere parametrized by $(\theta_1,\phi_1)$ remains of finite size near the apex of the conifold.

Let us now move from $T^{1,1}$ to $Y^{p,q}$. Calabi-Yau metrics describing small resolutions of the cone over $Y^{p,q}$ manifolds were found in \cite{Martelli:2007pv} (see also \cite{Martelli:2007mk}) representing the analogue of \eqref{resolvedconifold} with $C(T^{1,1})$ replaced by $C(Y^{p,q})$. Given what we have just reviewed, it is natural to wonder whether such metrics can also be found from the uplift of a solution to 8d gauged supergravity\footnote{The author thanks Dario Martelli for suggesting that this should be the case.}, and in this subsection we answer this question positively. We focus for simplicity on the case $\text{dim}(Y^{p,q})=5$, or Calabi-Yau three-folds. The metrics of \cite{Martelli:2007pv,Martelli:2007mk} involve a K\"ahler-Einstein metric of constant curvature, which like in Section \ref{sec:8dsugra} we take to be a two-sphere. Then, obtaining the expression of the 8d fields is straightforward if one considers the 11d metric describing M-theory on the small resolution of $C(Y^{p,q})$,
\es{CYpq_res}{
\ds_{11}&=\ds(\R^{1,4})+\ds(Y_6)\,,\\
\ds(Y_6)&=(y-x)\left(\frac{\diff x^2}{X}+\frac{\diff y^2}{Y}\right)+(y-1)(1-x)(\diff\theta^2+\sin^2\theta\,\diff\alpha^2)\\
&+\frac{X}{y-x}[\diff z+(y-1)(\diff\psi-\cos\theta\,\diff\alpha)]^2
+\frac{Y}{y-x}[\diff z+(x-1)(\diff\psi-\cos\theta\,\diff\alpha)]^2\,,
}
where
\es{XY_resolution}{
X&=x-1+\frac{2}{3}(x-1)^2+\frac{2\mu}{x-1}\,,\\
Y&=1-y-\frac{2}{3}(1-y)^2+\frac{2\nu}{y-1}\,.
}
We observe that this can also be found from the uplift of a solution of 8d supergravity, and in particular of the truncation presented in \eqref{8daction}. The expression for the corresponding 8d fields is
\es{CYpq_res_8d}{
\ds_8=&\ex^{\phi/3}\,\left[\ds(\R^{1,4})+(y-x)\left(\frac{\diff x^2}{X}+\frac{\diff y^2}{Y}+\frac{X\,Y}{Z}\diff z^2\right)\right]\,,\\
A=&\frac{X(1-y)+Y(1-x)}{Z}\frac{\diff z}{g}\,,\\
\ex^{2\phi}=&\frac{g^6(1-x)^2(y-1)^2\,Z}{y-x}\,,\quad
\ex^{6\lambda}=\frac{Z}{(1-x)(y-x)(y-1)}\,,
}
where the function $Z$ is given by
\es{Z8d}{
Z=(y-1)^2\,X+(1-x)^2\,Y\,.
}
In the limit $x\to \infty$ one recovers the solutions described in Section \ref{sec:8dsugra}, representing D6 branes wrapped on a spindle or 11d supergravity on the cone over $Y^{p,q}$ manifolds. For finite $x$ and $y$, the global analysis of this metric was carried out in \cite{Martelli:2007pv} and we refer the reader to that paper for details.

We note that while the authors of \cite{Martelli:2007pv,Martelli:2007mk} focused mostly on the case $n>0$, which leads to Calabi-Yau manifolds of real dimension six and higher, the case $n=0$ is also interesting and gives a Calabi-Yau two-fold with metric
\es{CY2}{
\ds_4=(x-y)\left(\frac{\diff x^2}{X}+\frac{\diff y^2}{Y}\right)+\frac{Y}{y-x}(\diff z+(1-x)\diff\psi)+\frac{X}{y-x}(\diff z+(1-y)\diff\psi)\,,
}
where 
\es{}{
X=\frac{x^2+4\mu-1}{2}\,,\quad
Y=-\frac{y^2+4\nu-1}{2}\,.
}
In the case where the twist condition is chosen for the bundle over the spindle, asymptotically (at $x=\infty$) this represents the non-compact orbifold $\mathcal{O}(-n_1-n_2)\to \spindle$, where $\spindle\equiv \mathbb{WCP}_{[n_1,n_2]}$. At finite $x$, \eqref{CY2} is a partial resolution of the conical singularity, where the apex of the cone is replaced with the spindle $\spindle$. This construction is analyized, from an abstract perspective, in Section 3.3.2 of \cite{Martelli:2023oqk}, while the same problem is studied with an explicit metric in \cite{DarioToAppear}, to which we refer for more details.\footnote{The author would like to thank Dario Martelli for clarifications on the comments made here on the $n=0$ case.}

\newpage

\section{Discussion}\label{sec:discussion}

In this paper we have considered solutions of 8d gauged supergravity whose uplift to type IIA describes a stack of D6 branes wrapped on a spindle $\spindle$. A further uplift gives 11d supergravity on a spacetime which is the product of $\R^{1,3}$ and the Calabi-Yau cone over five-dimensional $Y^{p,q}$ Sasaki-Einstein manifolds. This gives a new perspective on $Y^{p,q}$ manifolds, which we write as a circle fibration over the warped product between a spindle and a K\"ahler-Einstein manifold with a constant curvature metric. The spindle part of this fibration is found to always realize the twist condition \cite{Ferrero:2021etw}, except in the case of Sasaki-Einstein manifolds of dimension three, when both the twist and the anti-twist are possible. We also considered the small resolution of the cone over $Y^{p,q}$, for which a class of Calabi-Yau metrics was presented in \cite{Martelli:2007pv}: in that case, too, it is possible to obtain such metrics from the uplift of supersymmetric solutions of 8d gauged supergravity. 

The setup considered here is a special case of a family of solutions in various dimensions, which we present in \cite{Boisvert:2024jrl}, where we show that it is possible to find backreacted solutions representing branes wrapped on spindles even in cases where the field theory living on the worldvolume of the branes is not conformal. More generally, a lot of attention in recent years has been devoted to the study of supergravity backgrounds involving AdS factors, but it is quite possible that some of the constructions and techniques that have been investigated in that context apply to more general setups in which conformal invariance is not present, and it would be interesting to investigate this in the future. We refer the reader to \cite{Boisvert:2024jrl} for additional comments on this point.

So far as the specific solutions considered in this paper are concerned, an interesting question is whether the observations made here can be generalized and used to find new special-holonomy metrics using gauged supergravity, and/or some inputs from the literature on spindle solutions. In particular, it would be interesting to find Sasaki-Einstein metrics that arise from fibrations over four-dimensional orbifolds. Examples of these that have been recently introduced are fibrations of a spindle over another spindle \cite{Cheung:2022ilc,Bomans:2023ouw,Couzens:2022lvg} or the quadrilaterals of \cite{Faedo:2024upq}. Similarly, supersymmetric reductions K\"ahler metrics on $\mathbb{WCP}^2$ have already appeard in \cite{Gauntlett:2004zh} and, more recently, in \cite{Bianchi:2021uhn,Macpherson:2024frt}.

\section*{Acknowledgments}

The author would like to thank Jerome Gauntlett,Carlos Nunez and James Sparks for discussions. I am especially tankful to Dario Martelli for useful conversations and his extensive comments on this manuscript.

\appendix

\section{Gamma matrices}\label{app:gammas}

In this paper we use the symbol $\gamma$ to denote spacetime gamma matrices, {\it i.e.} satisfying a Clifford algebra in flat Lorentzian space with mostly plus signature. We use a superscript to denote the dimensionality of the underlying Lorentzian spacetime. In a similar spirit, we use the symbol $\tau$ to denote gamma matrices satisfying the Euclidean Clifford algebra, also indicating the corresponding dimension with a superscript. In both cases, for even dimensions the chirality matrix is denoted with a subscript $\star$. Contractions of differential forms with Clifford algebra elements are performed with unit weight, that is for a $p$-form $\omega_{(p)}$ we define
\es{}{
\slashed{\omega}=\omega_{\mu_1\ldots\mu_p}\gamma^{\mu_1\ldots \mu_p}\,.
}

\newpage

\providecommand{\href}[2]{#2}\begingroup\raggedright\endgroup


\begin{thebibliography}{10}

\bibitem{Sparks:2010sn}
J.~Sparks, ``{Sasaki-Einstein Manifolds},''
  \href{http://dx.doi.org/10.4310/SDG.2011.v16.n1.a6}{{\em Surveys Diff. Geom.}
  {\bfseries 16} (2011) 265--324},
  \href{http://arxiv.org/abs/1004.2461}{{\ttfamily arXiv:1004.2461 [math.DG]}}.

\bibitem{Gauntlett:2004yd}
J.~P. Gauntlett, D.~Martelli, J.~Sparks, and D.~Waldram, ``{Sasaki-Einstein
  metrics on S**2 x S**3},''
  \href{http://dx.doi.org/10.4310/ATMP.2004.v8.n4.a3}{{\em Adv. Theor. Math.
  Phys.} {\bfseries 8} no.~4, (2004) 711--734},
  \href{http://arxiv.org/abs/hep-th/0403002}{{\ttfamily arXiv:hep-th/0403002}}.

\bibitem{Gauntlett:2004hh}
J.~P. Gauntlett, D.~Martelli, J.~F. Sparks, and D.~Waldram, ``{A New infinite
  class of Sasaki-Einstein manifolds},''
  \href{http://dx.doi.org/10.4310/ATMP.2004.v8.n6.a3}{{\em Adv. Theor. Math.
  Phys.} {\bfseries 8} no.~6, (2004) 987--1000},
  \href{http://arxiv.org/abs/hep-th/0403038}{{\ttfamily arXiv:hep-th/0403038}}.

\bibitem{Gauntlett:2004zh}
J.~P. Gauntlett, D.~Martelli, J.~Sparks, and D.~Waldram, ``{Supersymmetric
  AdS(5) solutions of M theory},''
  \href{http://dx.doi.org/10.1088/0264-9381/21/18/005}{{\em Class. Quant.
  Grav.} {\bfseries 21} (2004) 4335--4366},
  \href{http://arxiv.org/abs/hep-th/0402153}{{\ttfamily arXiv:hep-th/0402153}}.

\bibitem{Cvetic:2005ft}
M.~Cvetic, H.~Lu, D.~N. Page, and C.~N. Pope, ``{New Einstein-Sasaki spaces in
  five and higher dimensions},''
  \href{http://dx.doi.org/10.1103/PhysRevLett.95.071101}{{\em Phys. Rev. Lett.}
  {\bfseries 95} (2005) 071101},
  \href{http://arxiv.org/abs/hep-th/0504225}{{\ttfamily arXiv:hep-th/0504225}}.

\bibitem{Cvetic:2005vk}
M.~Cvetic, H.~Lu, D.~N. Page, and C.~N. Pope, ``{New Einstein-Sasaki and
  Einstein spaces from Kerr-de Sitter},''
  \href{http://dx.doi.org/10.1088/1126-6708/2009/07/082}{{\em JHEP} {\bfseries
  07} (2009) 082}, \href{http://arxiv.org/abs/hep-th/0505223}{{\ttfamily
  arXiv:hep-th/0505223}}.

\bibitem{Friedrich:2001nh}
T.~Friedrich and S.~Ivanov, ``{Parallel spinors and connections with skew
  symmetric torsion in string theory},'' {\em Asian J. Math.} {\bfseries 6}
  (2002) 303--336, \href{http://arxiv.org/abs/math/0102142}{{\ttfamily
  arXiv:math/0102142}}.

\bibitem{Gauntlett:2001ur}
J.~P. Gauntlett, N.~Kim, D.~Martelli, and D.~Waldram, ``{Five-branes wrapped on
  SLAG three cycles and related geometry},''
  \href{http://dx.doi.org/10.1088/1126-6708/2001/11/018}{{\em JHEP} {\bfseries
  11} (2001) 018}, \href{http://arxiv.org/abs/hep-th/0110034}{{\ttfamily
  arXiv:hep-th/0110034}}.

\bibitem{Gauntlett:2002fz}
J.~P. Gauntlett and S.~Pakis, ``{The Geometry of D = 11 killing spinors},''
  \href{http://dx.doi.org/10.1088/1126-6708/2003/04/039}{{\em JHEP} {\bfseries
  04} (2003) 039}, \href{http://arxiv.org/abs/hep-th/0212008}{{\ttfamily
  arXiv:hep-th/0212008}}.

\bibitem{Gauntlett:2002sc}
J.~P. Gauntlett, D.~Martelli, S.~Pakis, and D.~Waldram, ``{G structures and
  wrapped NS5-branes},''
  \href{http://dx.doi.org/10.1007/s00220-004-1066-y}{{\em Commun. Math. Phys.}
  {\bfseries 247} (2004) 421--445},
  \href{http://arxiv.org/abs/hep-th/0205050}{{\ttfamily arXiv:hep-th/0205050}}.

\bibitem{Kim:2005ez}
N.~Kim, ``{AdS(3) solutions of IIB supergravity from D3-branes},''
  \href{http://dx.doi.org/10.1088/1126-6708/2006/01/094}{{\em JHEP} {\bfseries
  01} (2006) 094}, \href{http://arxiv.org/abs/hep-th/0511029}{{\ttfamily
  arXiv:hep-th/0511029}}.

\bibitem{Kim:2006qu}
N.~Kim and J.-D. Park, ``{Comments on AdS(2) solutions of D=11 supergravity},''
  \href{http://dx.doi.org/10.1088/1126-6708/2006/09/041}{{\em JHEP} {\bfseries
  09} (2006) 041}, \href{http://arxiv.org/abs/hep-th/0607093}{{\ttfamily
  arXiv:hep-th/0607093}}.

\bibitem{Gauntlett:2007ts}
J.~P. Gauntlett and N.~Kim, ``{Geometries with Killing Spinors and
  Supersymmetric AdS Solutions},''
  \href{http://dx.doi.org/10.1007/s00220-008-0575-5}{{\em Commun. Math. Phys.}
  {\bfseries 284} (2008) 897--918},
  \href{http://arxiv.org/abs/0710.2590}{{\ttfamily arXiv:0710.2590 [hep-th]}}.

\bibitem{Ferrero:2020laf}
P.~Ferrero, J.~P. Gauntlett, J.~M. P\'erez Ipi\~na, D.~Martelli, and J.~Sparks,
  ``{D3-Branes Wrapped on a Spindle},''
  \href{http://dx.doi.org/10.1103/PhysRevLett.126.111601}{{\em Phys. Rev.
  Lett.} {\bfseries 126} no.~11, (2021) 111601},
  \href{http://arxiv.org/abs/2011.10579}{{\ttfamily arXiv:2011.10579
  [hep-th]}}.

\bibitem{Hosseini:2021fge}
S.~M. Hosseini, K.~Hristov, and A.~Zaffaroni, ``{Rotating multi-charge spindles
  and their microstates},''
  \href{http://dx.doi.org/10.1007/JHEP07(2021)182}{{\em JHEP} {\bfseries 07}
  (2021) 182}, \href{http://arxiv.org/abs/2104.11249}{{\ttfamily
  arXiv:2104.11249 [hep-th]}}.

\bibitem{Boido:2021szx}
A.~Boido, J.~M.~P. Ipi\~na, and J.~Sparks, ``{Twisted D3-brane and M5-brane
  compactifications from multi-charge spindles},''
  \href{http://dx.doi.org/10.1007/JHEP07(2021)222}{{\em JHEP} {\bfseries 07}
  (2021) 222}, \href{http://arxiv.org/abs/2104.13287}{{\ttfamily
  arXiv:2104.13287 [hep-th]}}.

\bibitem{Ferrero:2021etw}
P.~Ferrero, J.~P. Gauntlett, and J.~Sparks, ``{Supersymmetric spindles},''
  \href{http://dx.doi.org/10.1007/JHEP01(2022)102}{{\em JHEP} {\bfseries 01}
  (2022) 102}, \href{http://arxiv.org/abs/2112.01543}{{\ttfamily
  arXiv:2112.01543 [hep-th]}}.

\bibitem{Ferrero:2020twa}
P.~Ferrero, J.~P. Gauntlett, J.~M.~P. Ipi\~na, D.~Martelli, and J.~Sparks,
  ``{Accelerating black holes and spinning spindles},''
  \href{http://dx.doi.org/10.1103/PhysRevD.104.046007}{{\em Phys. Rev. D}
  {\bfseries 104} no.~4, (2021) 046007},
  \href{http://arxiv.org/abs/2012.08530}{{\ttfamily arXiv:2012.08530
  [hep-th]}}.

\bibitem{Cassani:2021dwa}
D.~Cassani, J.~P. Gauntlett, D.~Martelli, and J.~Sparks, ``{Thermodynamics of
  accelerating and supersymmetric AdS4 black holes},''
  \href{http://dx.doi.org/10.1103/PhysRevD.104.086005}{{\em Phys. Rev. D}
  {\bfseries 104} no.~8, (2021) 086005},
  \href{http://arxiv.org/abs/2106.05571}{{\ttfamily arXiv:2106.05571
  [hep-th]}}.

\bibitem{Ferrero:2021ovq}
P.~Ferrero, M.~Inglese, D.~Martelli, and J.~Sparks, ``{Multicharge accelerating
  black holes and spinning spindles},''
  \href{http://dx.doi.org/10.1103/PhysRevD.105.126001}{{\em Phys. Rev. D}
  {\bfseries 105} no.~12, (2022) 126001},
  \href{http://arxiv.org/abs/2109.14625}{{\ttfamily arXiv:2109.14625
  [hep-th]}}.

\bibitem{Couzens:2021cpk}
C.~Couzens, ``{A tale of (M)2 twists},''
  \href{http://dx.doi.org/10.1007/JHEP03(2022)078}{{\em JHEP} {\bfseries 03}
  (2022) 078}, \href{http://arxiv.org/abs/2112.04462}{{\ttfamily
  arXiv:2112.04462 [hep-th]}}.

\bibitem{Ferrero:2021wvk}
P.~Ferrero, J.~P. Gauntlett, D.~Martelli, and J.~Sparks, ``{M5-branes wrapped
  on a spindle},'' \href{http://dx.doi.org/10.1007/JHEP11(2021)002}{{\em JHEP}
  {\bfseries 11} (2021) 002}, \href{http://arxiv.org/abs/2105.13344}{{\ttfamily
  arXiv:2105.13344 [hep-th]}}.

\bibitem{Giri:2021xta}
S.~Giri, ``{Black holes with spindles at the horizon},''
  \href{http://dx.doi.org/10.1007/JHEP06(2022)145}{{\em JHEP} {\bfseries 06}
  (2022) 145}, \href{http://arxiv.org/abs/2112.04431}{{\ttfamily
  arXiv:2112.04431 [hep-th]}}.

\bibitem{Faedo:2021nub}
F.~Faedo and D.~Martelli, ``{D4-branes wrapped on a spindle},''
  \href{http://dx.doi.org/10.1007/JHEP02(2022)101}{{\em JHEP} {\bfseries 02}
  (2022) 101}, \href{http://arxiv.org/abs/2111.13660}{{\ttfamily
  arXiv:2111.13660 [hep-th]}}.

\bibitem{Couzens:2022yiv}
C.~Couzens and K.~Stemerdink, ``{Universal spindles: D2's on $\Sigma$ and M5's
  on $\Sigma\times \mathbb{H}^3$},''
  \href{http://arxiv.org/abs/2207.06449}{{\ttfamily arXiv:2207.06449
  [hep-th]}}.

\bibitem{Couzens:2022aki}
C.~Couzens, N.~T. Macpherson, and A.~Passias, ``{A plethora of Type IIA
  embeddings for d = 5 minimal supergravity},''
  \href{http://dx.doi.org/10.1007/JHEP01(2023)047}{{\em JHEP} {\bfseries 01}
  (2023) 047}, \href{http://arxiv.org/abs/2209.15540}{{\ttfamily
  arXiv:2209.15540 [hep-th]}}.

\bibitem{Arav:2022lzo}
I.~Arav, J.~P. Gauntlett, M.~M. Roberts, and C.~Rosen, ``{Leigh-Strassler
  compactified on a spindle},''
  \href{http://dx.doi.org/10.1007/JHEP10(2022)067}{{\em JHEP} {\bfseries 10}
  (2022) 067}, \href{http://arxiv.org/abs/2207.06427}{{\ttfamily
  arXiv:2207.06427 [hep-th]}}.

\bibitem{Suh:2022pkg}
M.~Suh, ``{Spindle black holes from mass-deformed ABJM},''
  \href{http://arxiv.org/abs/2211.11782}{{\ttfamily arXiv:2211.11782
  [hep-th]}}.

\bibitem{Suh:2023xse}
M.~Suh, ``{Baryonic spindles from conifolds},''
  \href{http://arxiv.org/abs/2304.03308}{{\ttfamily arXiv:2304.03308
  [hep-th]}}.

\bibitem{Hristov:2023rel}
K.~Hristov and M.~Suh, ``{Spindle black holes in AdS$_{4} \times$ SE$_{7}$},''
  \href{http://dx.doi.org/10.1007/JHEP10(2023)141}{{\em JHEP} {\bfseries 10}
  (2023) 141}, \href{http://arxiv.org/abs/2307.10378}{{\ttfamily
  arXiv:2307.10378 [hep-th]}}.

\bibitem{Amariti:2023gcx}
A.~Amariti, S.~Mancani, D.~Morgante, N.~Petri, and A.~Segati, ``{BBBW on the
  spindle},'' \href{http://arxiv.org/abs/2309.11362}{{\ttfamily
  arXiv:2309.11362 [hep-th]}}.

\bibitem{Amariti:2023mpg}
A.~Amariti, N.~Petri, and A.~Segati, ``{T$^{1,1}$ truncation on the spindle},''
  \href{http://dx.doi.org/10.1007/JHEP07(2023)087}{{\em JHEP} {\bfseries 07}
  (2023) 087}, \href{http://arxiv.org/abs/2304.03663}{{\ttfamily
  arXiv:2304.03663 [hep-th]}}.

\bibitem{Bah:2021mzw}
I.~Bah, F.~Bonetti, R.~Minasian, and E.~Nardoni, ``{Holographic Duals of
  Argyres-Douglas Theories},''
  \href{http://dx.doi.org/10.1103/PhysRevLett.127.211601}{{\em Phys. Rev.
  Lett.} {\bfseries 127} no.~21, (2021) 211601},
  \href{http://arxiv.org/abs/2105.11567}{{\ttfamily arXiv:2105.11567
  [hep-th]}}.

\bibitem{Couzens:2021tnv}
C.~Couzens, N.~T. Macpherson, and A.~Passias, ``{$ \mathcal{N} $ = (2, 2)
  AdS$_{3}$ from D3-branes wrapped on Riemann surfaces},''
  \href{http://dx.doi.org/10.1007/JHEP02(2022)189}{{\em JHEP} {\bfseries 02}
  (2022) 189}, \href{http://arxiv.org/abs/2107.13562}{{\ttfamily
  arXiv:2107.13562 [hep-th]}}.

\bibitem{Suh:2021ifj}
M.~Suh, ``{D3-branes and M5-branes wrapped on a topological disc},''
  \href{http://dx.doi.org/10.1007/JHEP03(2022)043}{{\em JHEP} {\bfseries 03}
  (2022) 043}, \href{http://arxiv.org/abs/2108.01105}{{\ttfamily
  arXiv:2108.01105 [hep-th]}}.

\bibitem{Suh:2021hef}
M.~Suh, ``{M2-branes wrapped on a topological disk},''
  \href{http://dx.doi.org/10.1007/JHEP09(2022)048}{{\em JHEP} {\bfseries 09}
  (2022) 048}, \href{http://arxiv.org/abs/2109.13278}{{\ttfamily
  arXiv:2109.13278 [hep-th]}}.

\bibitem{Couzens:2021rlk}
C.~Couzens, K.~Stemerdink, and D.~van~de Heisteeg, ``{M2-branes on discs and
  multi-charged spindles},''
  \href{http://dx.doi.org/10.1007/JHEP04(2022)107}{{\em JHEP} {\bfseries 04}
  (2022) 107}, \href{http://arxiv.org/abs/2110.00571}{{\ttfamily
  arXiv:2110.00571 [hep-th]}}.

\bibitem{Suh:2021aik}
M.~Suh, ``{D4-branes wrapped on a topological disk},''
  \href{http://dx.doi.org/10.1007/JHEP06(2023)008}{{\em JHEP} {\bfseries 06}
  (2023) 008}, \href{http://arxiv.org/abs/2108.08326}{{\ttfamily
  arXiv:2108.08326 [hep-th]}}.

\bibitem{Bah:2021hei}
I.~Bah, F.~Bonetti, R.~Minasian, and E.~Nardoni, ``{M5-brane sources,
  holography, and Argyres-Douglas theories},''
  \href{http://dx.doi.org/10.1007/JHEP11(2021)140}{{\em JHEP} {\bfseries 11}
  (2021) 140}, \href{http://arxiv.org/abs/2106.01322}{{\ttfamily
  arXiv:2106.01322 [hep-th]}}.

\bibitem{Couzens:2022yjl}
C.~Couzens, H.~Kim, N.~Kim, and Y.~Lee, ``{Holographic duals of M5-branes on an
  irregularly punctured sphere},''
  \href{http://dx.doi.org/10.1007/JHEP07(2022)102}{{\em JHEP} {\bfseries 07}
  (2022) 102}, \href{http://arxiv.org/abs/2204.13537}{{\ttfamily
  arXiv:2204.13537 [hep-th]}}.

\bibitem{Karndumri:2022wpu}
P.~Karndumri and P.~Nuchino, ``{Five-branes wrapped on topological disks from
  7D N=2 gauged supergravity},''
  \href{http://dx.doi.org/10.1103/PhysRevD.105.066010}{{\em Phys. Rev. D}
  {\bfseries 105} no.~6, (2022) 066010},
  \href{http://arxiv.org/abs/2201.05037}{{\ttfamily arXiv:2201.05037
  [hep-th]}}.

\bibitem{Couzens:2023kyf}
C.~Couzens, M.~J. Kang, C.~Lawrie, and Y.~Lee, ``{Holographic duals of Higgsed
  $\mathcal{D}_p^b(BCD)$},'' \href{http://arxiv.org/abs/2312.12503}{{\ttfamily
  arXiv:2312.12503 [hep-th]}}.

\bibitem{Faedo:2022rqx}
F.~Faedo, A.~Fontanarossa, and D.~Martelli, ``{Branes wrapped on orbifolds and
  their gravitational blocks},''
  \href{http://dx.doi.org/10.1007/s11005-023-01671-1}{{\em Lett. Math. Phys.}
  {\bfseries 113} no.~3, (2023) 51},
  \href{http://arxiv.org/abs/2210.16128}{{\ttfamily arXiv:2210.16128
  [hep-th]}}.

\bibitem{Suh:2022olh}
M.~Suh, ``{M5-branes and D4-branes wrapped on a direct product of spindle and
  Riemann surface},'' \href{http://arxiv.org/abs/2207.00034}{{\ttfamily
  arXiv:2207.00034 [hep-th]}}.

\bibitem{Couzens:2022lvg}
C.~Couzens, H.~Kim, N.~Kim, Y.~Lee, and M.~Suh, ``{D4-branes wrapped on
  four-dimensional orbifolds through consistent truncation},''
  \href{http://dx.doi.org/10.1007/JHEP02(2023)025}{{\em JHEP} {\bfseries 02}
  (2023) 025}, \href{http://arxiv.org/abs/2210.15695}{{\ttfamily
  arXiv:2210.15695 [hep-th]}}.

\bibitem{Cheung:2022ilc}
K.~C.~M. Cheung, J.~H.~T. Fry, J.~P. Gauntlett, and J.~Sparks, ``{M5-branes
  wrapped on four-dimensional orbifolds},''
  \href{http://dx.doi.org/10.1007/JHEP08(2022)082}{{\em JHEP} {\bfseries 08}
  (2022) 082}, \href{http://arxiv.org/abs/2204.02990}{{\ttfamily
  arXiv:2204.02990 [hep-th]}}.

\bibitem{Bomans:2023ouw}
P.~Bomans, C.~Couzens, Y.~Lee, and S.~Ning, ``{Symmetry Breaking and Consistent
  Truncations from M5-branes Wrapping a Disc},''
  \href{http://arxiv.org/abs/2308.08616}{{\ttfamily arXiv:2308.08616
  [hep-th]}}.

\bibitem{Faedo:2024upq}
F.~Faedo, A.~Fontanarossa, and D.~Martelli, ``{Branes wrapped on
  quadrilaterals},'' \href{http://arxiv.org/abs/2402.08724}{{\ttfamily
  arXiv:2402.08724 [hep-th]}}.

\bibitem{Macpherson:2024frt}
N.~T. Macpherson, P.~Merrikin, and C.~Nunez, ``{Marginally deformed
  AdS$_5$/CFT$_4$ and spindle-like orbifolds},''
  \href{http://arxiv.org/abs/2403.02380}{{\ttfamily arXiv:2403.02380
  [hep-th]}}.

\bibitem{Boisvert:2024jrl}
M.~Boisvert and P.~Ferrero, ``{A story of non-conformal branes: spindles,
  disks, circles and black holes},''
  \href{http://arxiv.org/abs/2403.03989}{{\ttfamily arXiv:2403.03989
  [hep-th]}}.

\bibitem{Edelstein:2001pu}
J.~D. Edelstein and C.~Nunez, ``{D6-branes and M theory geometrical transitions
  from gauged supergravity},''
  \href{http://dx.doi.org/10.1088/1126-6708/2001/04/028}{{\em JHEP} {\bfseries
  04} (2001) 028}, \href{http://arxiv.org/abs/hep-th/0103167}{{\ttfamily
  arXiv:hep-th/0103167}}.

\bibitem{Boyer:1999ms}
C.~P. Boyer and K.~Galicki, ``{On Sasakian-Einstein geometry},''
  \href{http://dx.doi.org/10.1142/S0129167X00000477}{{\em Int. J. Math.}
  {\bfseries 11} (2000) 873--909},
  \href{http://arxiv.org/abs/math/9811098}{{\ttfamily arXiv:math/9811098}}.

\bibitem{Witten:1988ze}
E.~Witten, ``{Topological Quantum Field Theory},''
  \href{http://dx.doi.org/10.1007/BF01223371}{{\em Commun. Math. Phys.}
  {\bfseries 117} (1988) 353}.

\bibitem{Bianchi:2021uhn}
M.~Bianchi, U.~Bruzzo, P.~Fr\'e, and D.~Martelli, ``{Resolution \`a la
  Kronheimer of $\mathbb {C}^3/\Gamma $ singularities and the
  Monge\textendash{}Amp\`ere equation for Ricci-flat K\"ahler metrics in view
  of D3-brane solutions of supergravity},''
  \href{http://dx.doi.org/10.1007/s11005-021-01420-2}{{\em Lett. Math. Phys.}
  {\bfseries 111} no.~3, (2021) 79},
  \href{http://arxiv.org/abs/2105.11704}{{\ttfamily arXiv:2105.11704
  [math.DG]}}.

\bibitem{Martelli:2023oqk}
D.~Martelli and A.~Zaffaroni, ``{Equivariant localization and holography},''
  \href{http://arxiv.org/abs/2306.03891}{{\ttfamily arXiv:2306.03891
  [hep-th]}}.

\bibitem{Inglese:2023tyc}
M.~Inglese, D.~Martelli, and A.~Pittelli, ``{Supersymmetry and Localization on
  Three-Dimensional Orbifolds},''
  \href{http://arxiv.org/abs/2312.17086}{{\ttfamily arXiv:2312.17086
  [hep-th]}}.

\bibitem{Martelli:2007pv}
D.~Martelli and J.~Sparks, ``{Resolutions of non-regular Ricci-flat Kahler
  cones},'' \href{http://dx.doi.org/10.1016/j.geomphys.2009.06.005}{{\em J.
  Geom. Phys.} {\bfseries 59} (2009) 1175--1195},
  \href{http://arxiv.org/abs/0707.1674}{{\ttfamily arXiv:0707.1674 [math.DG]}}.

\bibitem{Martelli:2007mk}
D.~Martelli and J.~Sparks, ``{Baryonic branches and resolutions of Ricci-flat
  Kahler cones},'' \href{http://dx.doi.org/10.1088/1126-6708/2008/04/067}{{\em
  JHEP} {\bfseries 04} (2008) 067},
  \href{http://arxiv.org/abs/0709.2894}{{\ttfamily arXiv:0709.2894 [hep-th]}}.

\bibitem{Salam:1984ft}
A.~Salam and E.~Sezgin, ``{d=8 supergravity},''
  \href{http://dx.doi.org/10.1016/0550-3213(85)90613-3}{{\em Nucl. Phys. B}
  {\bfseries 258} (1985) 284--304}.

\bibitem{DarioToAppear}
M.~K. Crisafio, A.~Fontanarossa, and D.~Martelli, ``to appear,''.

\end{thebibliography}
\end{document}